\documentclass[rmp,eqsecnum,12pt,onecolumn,a4paper,notitlepage,nofootinbib,longbibliography]{revtex4-1}
\pdfoutput=1
\usepackage[utf8]{inputenc}
\usepackage{hyperref}
\usepackage{amsfonts, amssymb, amsmath,amsthm}
\usepackage{graphicx}
\usepackage{nicefrac,amscd}
\usepackage{bbm,pifont}
\usepackage[all,pdf]{xy}
\usepackage{bm} 
\usepackage{physics} 
\usepackage{xcolor,enumitem}
\usepackage{fp}
\newcommand{\circled}[1]{\FPeval{\result}{clip(201+#1)}\text{\ding{\result}}} 

\newtheorem{thm}[equation]{Theorem}

\newtheorem{assumption}[equation]{Assumption}

\newtheorem{fact}[equation]{Fact}
\theoremstyle{definition}
\newtheorem{definition}[equation]{Definition}
\theoremstyle{remark}

\DeclareMathOperator{\myre}{Re}
\renewcommand{\Re}{\myre}

\begin{document}

\title{Pitowsky's Kolmogorovian models and Super-Determinism}
\date{\today}

\author{Jakob \surname{Kellner}}
\thanks{This paper is an extended version of my bachelor  thesis at the University of Vienna, supervised by Caslav Brukner. I would like to express my gratitude for his time and advice. Supported by Austrian Science Fund (FWF): P23875 and P26737, and FWF Individual Project (No.\ 24621)}
\email{Jakob.Kellner@TUWien.ac.at}
\homepage{http://dmg.tuwien.ac.at/kellner/}
\affiliation{Institute of Discrete Mathematics and Geometry, TU Wien, Austria}

\begin{abstract}
In an attempt to demonstrate that local hidden variables 
are mathematically possible,
Pitowsky 
constructed ``spin-$\nicefrac12$ functions''  and later 
``Kolmogorovian models'', which employs a nonstandard 
notion of probability.

We describe Pitowsky's analysis and 
argue (with the benefit of hindsight)
that his notion of hidden variables 
is in fact just super-determinism (and accordingly 
physically not relevant).

Pitowsky's first construction uses the Con\-ti\-nu\-um Hypothesis.
Farah and Magidor 
took this as an indication that
at some stage physics might give arguments
for or against adopting specific new axioms of set theory.
We would rather argue that it 
supports the opposing view, i.e., the widespread intuition ``if you need a non-measurable 
function, it is physically irrelevant''.

\end{abstract}
\maketitle

\section{No-go theorems}

We briefly recall the notion on hidden variables and the two no-go theorems 
that will be relevant in this paper: 
Bell's theorem~\citep{Bell1964},
the groundbreaking first proof that local hidden variables are impossible; and the
GHZ theorem~\citep{GHZ}.
The GHZ theorem is simpler and stronger, as it
shows that local hidden variables cannot be consistently assigned to a single system (of three particles); whereas Bell's theorem shows 
that certain statistical frequencies cannot be 
reproduced by local hidden variables.

More details can be found in, e.g., \citep{Mermin1993}.

We will only consider systems of one, two or three
spin-$\nicefrac12$ particles. $\sigma_{\vb*a}$ denotes the spin observable in direction $\vb*a$ (with possible values $\pm1$); 
if we are dealing with more than one particle,
the observable for the $i$-th particle is called $\sigma^i_{\vb*a}$.
We will only investigate idealized \emph{Gedanken}experiments and do not care how they could be implemented in reality.

\subsection{Hidden variables, super-determinism}\label{sec:blah}

\subsubsection*{Local hidden variables}
In quantum mechanics, a (pure) system 
is described by its state, a vector $\ket\Psi$ in some Hilbert space $\mathcal H$. For our purposes we can assume $\mathcal H$ to be finite dimensional, which somewhat simplifies notation.
For a given state and an 
observable $A$, the result of the measurement is generally not determined; we just get 
probabilities for certain outcomes.

It is natural to ask whether there is a description of
the system that provides deterministic predictions.
Let us call such a description ``hidden variables'':
A system  in hidden variable state $v$ has predetermined results 
$v(A)$ for all\footnote{or just: sufficiently many; finitely many is enough for our purposes}
observables $A$.
(In particular we require predictions for observables $A$, $B$ which do not commute; i.e., which cannot be measured simultaneously.)
 
Once we perform a measurement for $A$ (resulting in
$v(A)$) then the system will change into a new hidden
variable state $v'$; and if we then perform a
measurement for $B$ we get the result $v'(B)$.
Generally there is no reason to assume that $v(B)=v'(B)$. Actually, it is obvious that for non-commuting 
observables, the hidden variable \emph{has} to change:
Measuring first $A$, then $B$, and then $A$ again
will generally result in different values for
the two $A$ measurements.

Let us call hidden variables \emph{non-contextual},
if $v(B)=v'(B)$ \emph{is} satisfied
for \emph{commuting} observables $A$, $B$; and \emph{local}, if
it is satisfied for 
 \emph{spatially  
separated} observables.\footnote{Note that non-contextual implies local, as 
spatially separated observables have to commute (to prevent superluminal communication).}
In other words: if for a hidden variable state we ``simultaneously'' measure such
$A$, $B$ we 
get the results $v(A)$ and $v(B)$.

It is widely accepted that 
the GHZ theorem (cf. Section~\ref{sec:ghz}) shows that local hidden variables 
are impossible (assuming of course that quantum mechanical predictions are satisfied for all possible measurement combinations). 

Pitowsky claims his model is even non-contextual.

\subsubsection*{Statistical hidden variables}

A hidden variable model for a given quantum mechanical state $\ket\Psi$ must  
predict the results that are \emph{guaranteed} by quantum mechanics.\footnote{I.e.: If $\ket\Psi$ is eigenvector of $A$ with eigenvalue $a$, the we require $v(A)=a$.
Note that the GHZ theorem shows that not even that can be done with local hidden variables.}
But more generally,
it should also for other observables reproduce the predicted frequencies 
(i.e., frequencies other than 100\% or 0\%).
So we 
cannot assign the same hidden variable to all systems in state $\ket\Psi$ (as
there will be different results when
measuring some $A$ on different copies of $\ket\Psi$.)

So it is natural to assume that a certain classical probabilistic
mixture of different 
hidden variables represents $\ket\Psi$,
and that we can represent a sequence of systems in
state $\ket\Psi$ by 
``randomly'' (i.e., according to the measure)
picking hidden variables; and we require
that the resulting frequencies are those
predicted by
quantum mechanics.

More formally, we can require the following:
\begin{assumption}\label{asm:stat}
$\rho$ is  a probability measure 
on the set of hidden variables. 
For all observables $A$ (that we consider),
$\rho(\{v:\, v(A)=a\})$ has to be equal to the quantum mechanical probability to get result $a$ when measuring $A$.
\end{assumption}
(In Pitowsky's model, a weaker notion $\rho$
will be used instead of the classical probability 
measure.)

It is widely accepted that
Bell's theorem (cf.~Section~\ref{sec:bell})
shows that 
local statistical hidden variables  are impossible.

\emph{Non-local} statistical hidden variables
\emph{are} possible: A very simple toy model
(albeit with rather weird and unpleasant properties)
is given by \citet{kochenspecker}; a more serious theory is by \citet{bohm1,*bohm2}.

\subsubsection*{Super-determinism} 
 
Let us define as \emph{super-determinism} the statement:
\begin{quote}
All future phenomena  are determined by the present 
state (not just the
measurement results, but also the question which measurements are performed). 
\end{quote}
While this position might be philosophically 
reasonable or satisfying, it is 
useless for physics:
It is clear that there cannot ever be a feasible
``universal theory'' that predicts which measurements
will be performed. (We can make the measurements, i.e.,
the setting of some detector, depend 
on the arrival of photons from distance galaxies, etc.)

From a super-deterministic point of view,
hidden variables are irrelevant but possible:
we know which measurements will be performed,
and we can (but it makes little sense to do so) 
assign any values we like
to other measurements, and there is no reason 
to assume that these bogus values should satisfy any
quantum mechanical prediction.

Of course we can never prove that there are no
hidden variables of this ``perverted'' kind,
as we cannot (for obvious reasons) disprove super-determinism.

Non-super-deterministic
hidden variables should have the property that they
predict reasonable outcomes (i.e., outcomes compatible
with quantum mechanics)
for \emph{all} possible measurements (and not just for some 
measurements that specifically fit the hidden
variable). I.e., the hidden variable should not determine or restrict which measurement we are allowed
to perform.

\subsection{The Greenberger-Horne-Zeilinger (GHZ)  theorem}\label{sec:ghz}
\begin{figure}
\caption{\label{fig:pentagram1}The GHZ pentagram.}
\scalebox{.8}{$
\begin{xy}
0;<7cm,0cm>:
(0,0)*+{\sigma^1_y}="11"; 
(-0.5,-0.363271)*+{\sigma^1_x\sigma^2_x\sigma^3_x}="21";
(-0.118034,-0.363271)*+{\sigma^1_y\sigma^2_y\sigma^3_x}="22";
(0.118034,-0.363271)*+{\sigma^1_y\sigma^2_x\sigma^3_y}="23";
(0.5,-0.363271)*+{\sigma^1_x\sigma^2_y\sigma^3_y}="24";
(-0.5,-0.363271)+(-0.1,0.08)*+{\text{\footnotesize ${+}1$}};
(-0.118034,-0.363271)+(-0.1,0.08)*+{\text{\footnotesize ${+}1$}};
(0.118034,-0.363271)+(0.1,0.08)*+{\text{\footnotesize ${+}1$}};
(0.5,-0.363271)+(0.1,0.08)*+{\text{\footnotesize ${-}1$}};
(-0.190983,-0.587785)*+{\sigma^3_x}="31";
(0.190983,-0.587785)*+{\sigma^3_y}="32";
(0,-0.726542)*+{\sigma^1_x}="41";
(-0.309017,-0.951056)*+{\sigma^2_y}="51";
(0.309017,-0.951056)*+{\sigma^2_x}="52";
(-0.654021,-1.2)*+{\text{\footnotesize $\circled1\ {\prod}{=}1$}}="l1";
(-0.389904, -1.2)*+{\text{\footnotesize $\circled2\ {\prod}{=}1$}}="l2";
(0.389904, -1.2)*+{\text{\footnotesize ${\circled3\ \prod}{=}1$}}="l3";
(0.654021,-1.2)*+{\text{\footnotesize ${\circled4\ \prod}{=}1$}}="l4";
(0.82,-0.363271)*+{\text{\footnotesize ${\prod}{=}{-1}$}}="l5";
{\ar@{.>} "51";"l1"};
{\ar@{.>} "51";"l2"};
{\ar@{.>} "52";"l3"};
{\ar@{.>} "52";"l4"};
{\ar@{.>} "24";"l5"};
{\ar@{--} "11";"22"};
{\ar@{--} "11";"23"};
{\ar@{--} "21";"22"};
{\ar@{--} "22";"23"};
{\ar@{--} "23";"24"};
{\ar@{--} "22";"31"};
{\ar@{--} "23";"32"};
{\ar@{--} "21";"31"};
{\ar@{--} "24";"32"};
{\ar@{--} "31";"41"};
{\ar@{--} "32";"41"};
{\ar@{--} "31";"51"};
{\ar@{--} "32";"52"};
{\ar@{--} "41";"51"};
{\ar@{--} "41";"52"};
\end{xy}
$}
\end{figure}
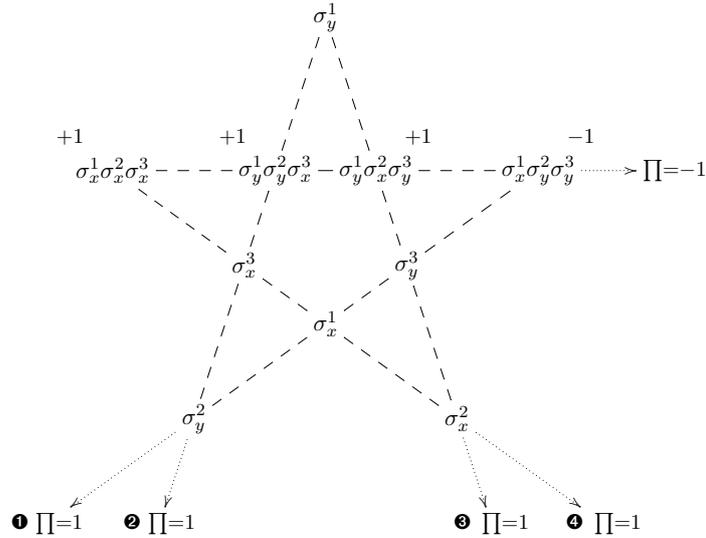

Consider a system of three spin-$\nicefrac12$ particles, and the operators listed in Figure~\ref{fig:pentagram1}. For each of the four lines
$\circled1$ to $\circled4$, the operators in the line commute and have product $+1$. 
Also, the four operators in the horizontal line commute and have product $-1$.
We cannot assign real numbers $v(A)$ to the operators while satisfying  these five 
requirements.\footnote{The ``requirement corresponding to
$\circled2$'' is 
$v(\sigma^1_y)\cdot v(\sigma^1_y\sigma^2_y\sigma^3_x)\cdot v(\sigma^3_x)\cdot v(\sigma^2_y)=1$, etc.}
(To see this, note that 
each node appears in exactly two lines.
So the product over the ``line products'', which 
is $1^4\cdot (-1)=-1$, has to be 
the product of all $v(A)$
squared, a contradiction.)

We now prepare the  system in 
a simultaneous eigenstate (the so-called 
GHZ state) for the operators in the 
horizontal line, with eigenvalue $-1$ for 
${\sigma^1_x\sigma^2_y\sigma^3_y}$ and $+1$ for the rest.
For each of the lines $\circled1$ to $\circled4$, 
call the product of the three remaining 
(single spin) operators  
the ``reduced product''.
E.g., the the reduced product of $\circled1$ is the product of $\sigma^1_x$, $\sigma^2_y$ and $\sigma^3_y$. 
So for the GHZ state, quantum mechanics predicts
the value $-1$ for the reduced product of
$\circled1$, and $+1$
for $\circled2$--$\circled4$.

Also, the single spin
%
operators are ``spatially
separated'', in the following sense:
We can choose which line from 
$\circled1$ to $\circled4$ we measure by 
choosing directions ($x$ or $y$) for each
particle. For example, $xyy$ results in $\circled1$ and $yyx$ in $\circled2$. So if we assume
that $v$ is a \emph{local} hidden variable, 
the result of measuring 
$\sigma^2_y$ will be $v(\sigma^2_y)$, irrespective
of whether we also measure $\sigma^1_y$ and $\sigma^3_x$ (i.e., we measure along $\circled2$) or whether we measure $\sigma^1_x$ and $\sigma^3_y$
(according to $\circled1$).

Therefore, any local hidden variable will
violate the quantum mechanically predicted 
value for the reduced product 
of at least one of the lines $\circled1$
to $\circled4$. 

For later reference, let us rephrase this result
using  projection operators: 
\begin{fact}\label{thm:GHZ}
We cannot assign ``yes'' or ''no'' to
the six tests ``Is $\sigma^i_{\vb*a}=+1$?''
such that all four of the following
requirements are met (which all follow from quantum mechanics for the GHZ state):
\begin{equation*}
\renewcommand{\arraystretch}{1.2}
\def\msp{{\phantom{.}}}
\left.
\begin{array}{ll}
\circled1&\text{an even}\\
\circled2&\text{an odd}\\
\circled3&\text{an odd}\\
\circled4&\text{an odd}
\end{array}\right\}
\text{number of testing}
\left\{
\begin{array}{ccc}
\sigma^1_x,\ &\sigma^2_y,\ &\sigma^3_y\\ 
\sigma^1_y,\ &\sigma^2_y,\ &\sigma^3_x  \\
\sigma^1_y,\ &\sigma^2_x,\ &\sigma^3_y  \\
\sigma^1_x,\ &\sigma^2_x,\ &\sigma^3_x
\end{array}
\right\}\text{for $+1$ results in ``yes''.}
\end{equation*}
\end{fact}

Of course, for any given system we can only test
one of the requirements $\circled1$ to $\circled4$.


\subsection{Bell's theorem}\label{sec:bell}


We now consider a pair of spin-$\nicefrac12$ particles 
in the singlet state $\frac1{\sqrt2}(\ket{\uparrow\downarrow}+\ket{\downarrow\uparrow})$.


 


For the singlet state, 
the probability 
to get the same result for $\sigma^1_{\vb*i}$ and $\sigma^2_{\vb*j}$ is
$p_{\vb*i,\vb*j}=\sin^2(\nicefrac\theta2)$,
where $\theta$ is the angle between $\vb*i$ and $\vb*j$.
Fix three directions  $\vb*a,\vb*b,\vb*c$ in the plane with angles of 120$^\circ$ between each two.
So $p_{\vb*a,\vb*b}=p_{\vb*a,\vb*c}=p_{\vb*b,\vb*c}=\nicefrac34$. 

We now consider a hidden variable,
i.e., a function $v$
that maps $\sigma^k_{\vb*i}$
to $v(\sigma^k_{\vb*i})=\pm 1$
for $k\in\{1,2\}$ and $\vb*i\in\{\vb*a,\vb*b,\vb*c\}$.

If we assume that the hidden variable is local
(and satisfies quantum mechanical predictions), we get
\begin{equation}\tag{$\ast$}\label{eq:huhu}
v(\sigma^2_{\vb*i})=-v(\sigma^1_{\vb*i}).
\end{equation}

Not all three of the numbers $v(\sigma^1_{\vb*a})$,
$v(\sigma^1_{\vb*b})$ and
$v(\sigma^1_{\vb*c})$ can be pairwise different.
So if \eqref{eq:huhu} holds, 
then there is at least one pair $\vb*i\neq \vb*j$
(call it ``chosen pair'') such that
$v(\sigma^1_{\vb*i})\neq v(\sigma^2_{\vb*j})$ (an event with quantum mechanical probability $\nicefrac14$).

Now we consider a (finite or infinite)
sequence of pairs in the singlet state; and assume
that the $n$-th pair is in some hidden variable state
$v_n$.
Assume that \eqref{eq:huhu} holds
(for each
$\vb*i\in\{\vb*a,\vb*b,\vb*c\}$)
with frequency at least $1-\epsilon$.
So with frequency $\ge 1-3\epsilon$, \eqref{eq:huhu} holds 
for all $\vb*i$ simultaneously, and then
there is at 
at least one ``chosen pair''. And as 
there are just three possible pairs,
at least one pair 
$\vb*i\neq \vb*j$ 
has to be chosen with frequency at least $\nicefrac13$
within the variables satisfying~\eqref{eq:huhu}; i.e., with frequency
at least $\nicefrac13-\epsilon$ within all variables.
When we chose $\epsilon$ to be $0.04$, say,
this shows:
\begin{fact}\label{thm:bell1}
It is not possible to find a (finite or infinite)
sequence of hidden
variables\footnote{Here, $v_n$ can be any
assignment from the set $\{\sigma^k_{\vb*i}:\, k=1,2,\ \vb*i=\vb*a,\vb*b,\vb*c
\}$ to $\pm1$}
$v_n$ such that 
the frequency\footnote{In case of infinite sequences we might require the frequency (i.e., the limit) to exist.} $f_{\vb*i,\vb*j}$ of
\begin{equation}\label{eq:bla1}
v_n(\sigma^1_{\vb*i})=v_n(\sigma^2_{\vb*j})
\end{equation}
is, for all 
$\vb*i,\vb*j\in\{\vb*a,\vb*b,\vb*c\}$,
within an error\footnote{I.e., $\abs{f_{\vb*i,\vb*j}-f^*_{\vb*i,\vb*j}}<0.04$.} of at most
$4\%$, equal to the 
quantum mechanical
predicted frequencies\footnote{which are 
$\nicefrac34$ for $\vb*i\neq \vb*j$ and $0$ for $\vb*i=\vb*j$}
$f^*_{\vb*i,\vb*j}$ 
for 
\begin{equation}\label{eq:bla2}
 \sigma^1_{\vb*i}\text{ has the same result as }\sigma^2_{\vb*j}.
 \end{equation}
\end{fact}

Assuming \emph{local} hidden variables,~\eqref{eq:bla1} is the same as~\eqref{eq:bla2}
(for each system in our sequence).

\section{Pitowsky's models}

In a series of articles, Pitowsky tried to analyze whether 
one could escape Bell's theorem and get \emph{local} models by using non-measurable functions. 
We  investigate the
following articles:
The  first attempt~\citep{pitowsky1982},
where he uses the Continuum Hypothesis  to construct 
a ``spin-$\nicefrac12$ function'' and a model for the singlet state. 
This attempt was immediately criticised~\citep{merminattacks,mcdonaldattacks}.
Starting with his response
\citeyearpar{pitowskyresponds}, Pitowsky formulated the idea
of using a nonstandard notion of probability,
culminating in the so-called 
\emph{Kolmogorovian model}~\citep[Section~5]{pitowsky1989} (in a construction which does 
not require the Continuum Hypothesis). This 
is a universal model that works for all quantum mechanical systems.

\subsection{Pitowsky's first attempt: The singlet state}\label{sec:model1}

To understand Pitowsky's analysis of Bell's theorem,
let us first give a consequence of this theorem.

\subsubsection*{Spin $\nicefrac12$ functions}

A ``spin $\nicefrac12$ function'' is a function 
$s_0$ from the sphere $S^2$ to $\pm1$ satisfying the following:
\begin{itemize}
  \item $s_0(-\vb*x)=-s_0(\vb*x)$.
  \item Fix $\vb*x$ and $0<\theta<\pi$ and set $S=\{\vb*y:\, \vb*y\cdot\vb*x=\cos(\theta) \}$
     (a circle equipped with the usual Lebesgue measure, with total measure $2\pi\sin(\theta)$).
     Then the set $S\cap \{\vb*y:\, s_0(\vb*y)=s_0(\vb*x)\}$ 
     is Lebesgue measurable in $S$ with relative measure $\cos^2(\nicefrac\theta2)$. 
\end{itemize}

Such a function looks promising for constructing hidden variables for the singlet state:
We can define the ``set of hidden variables''
to be the orthogonal group $\text{O}(3)$, equipped with
the normalized Haar measure $\theta$.
We then define 
for the hidden variable $g\in\text{O}(3)$
the measurement values
$g(\sigma^1_{\vb*a})=s_0(g(\vb*a))$ and 
$g(\sigma^2_{\vb*b})=-s_0(g(\vb*b))$.

If $s_0$ was additionally Lebesgue measurable,
then these hidden variables would actually work
and violate Bell's theorem.
So there cannot be a Lebesgue measurable spin $\nicefrac12$ function.\footnote{\label{fn:nine}In more detail:
Assume towards a contradiction that $s_0$ is a Lebesgue measurable spin $\nicefrac12$ function.

We now calculate the probability $p_{\vb*i,\vb*j}$ for
the event
$g(\sigma^1_{\vb*i})=g(\sigma^2_{\vb*j})$,
i.e., for
$s_0(g(\vb*i))\neq s_0(g(\vb*j))$, i.e.
for the two measurements having different outcome.
A ``random'' choice of $g$ corresponds to first
``randomly'' choosing 
a direction $\vb*i'$ (which is $g(\vb*i)$), and
then, again
``randomly'', another direction $\vb* j'$ 
such that $\vb*i'$ and $\vb* j'$
have the same angle $\theta$ as 
$\vb*i$ and $\vb* j$ (this $\vb*j$ is $g(\vb*j)$; note that $g$ preserves
angles). The probability that 
$s_0(\vb*j')=s_0(\vb*i')$ is, by the assumption,
$\cos^2(\nicefrac\theta2)$, so $p_{\vb*a,\vb*b}$
is $\sin^2(\nicefrac\theta2)$ as predicted by
quantum mechanics.
(Formally, we have to calculate
$\theta(X)$,
the $\text{O}(3)$ Haar measure 
of the set
$X=\{g\in \text{O}(3):\, s_0(g(\vb*a))\neq s_0(g(\vb*b))\}$.)

So we can, e.g., pick randomly an infinite (or: sufficiently large) sequence of hidden variables.
By the law of large numbers,
with probability one (or: sufficiently large,
nonzero is enough) we will get the correct
frequencies (or: frequencies that are sufficiently
close), contradicting 
Bell's theorem.} However, \citet{pitowsky1982} shows:
\begin{thm}
Assuming the Continuum Hypothesis, there is a (non measurable) spin-$\nicefrac12$ function. 
\end{thm}

Pitowsky then claims that the local hidden variables
for the singlet state 
(defined as above),
work and do not suffer from Bell's theorem.

It is not entirely clear to us from the 
paper~\cite{pitowsky1982} how he thought that
this would actually work out. 
Apparently he assumed that
by ``sabotaging'' Bell's proof
(by using non-measurable functions), and 
maybe using an intuition similar to the one outlined in footnote~\ref{fn:nine},
the problem would go away.

\subsubsection*{Criticism}\label{sec:criticism}

But of course it does not. This
was pointed out 
immediately~\citep{merminattacks,mcdonaldattacks}:

Let us paraphrase the criticism in concrete terms
(we use the notation of Section~\ref{sec:bell}):
Let us fix for example $N=50\,000$, and generate $N$
many singlet states.
Quantum mechanics tells us that 
for each pair $\vb*i,\vb*j$ from
$\{\vb*a,\vb*b,\vb*c\}$
the  frequency of the event
``$
\sigma^1_{\vb*i}\text{ equals } \sigma^2_{\vb* j}
$''
will  (virtually certainly)  be very close to the calculated probability of $\nicefrac34$.
More concretely, with probability $>1-10^{-91}$
(i.e., ``always'')
the frequency will be greater than $f_0=\nicefrac34-0.04$.
But Bell's theorem shows that
\emph{any} sequence of hidden variables,
\emph{irrespective of whether they were created by 
a classical random process, or some nonstandard method}
either gives a frequency $<f_0$ for at least 
one such pair  $\vb*i,\vb*j$,
or gives a frequency of $>0.04$ to the
(quantum mechanically outright impossible) event
``$
\sigma^1_{\vb*i}\text{ equals } \sigma^2_{\vb* i}
$''
for some $\vb*i$.

\subsubsection*{Response, towards the universal model}
In response,
\citet{pitowskyresponds} indicated that 
the objections are based on the ``classical'' notion of
probability, which has to be modified.
We need a new notion of probability (let us call it Pitowsky probability). In this notion, it can
happen that two sets (or: events) $A$, $B$  have Pitowsky probability one, but the intersection $A\cap B$ has ``classical'' probability zero.

Actually, it turns out that the intersection
even has to have small Pitowsky probability,
as \citet[p.~164]{pitowsky1983} elaborates:
According to his new notion of probability, it is no problem
if ``90\% all of the objects are red'' and 
``90\% all of the objects are small'' but
``no object is small and red''.\footnote{
We will later even see 
the following form of the effect:  
``All objects are red'' and
``all objects are small'' but 
``no object is small and red''.
This form was pointed out by~\citet[p. 1331]{jozsa},
who quite poignantly described
the ``paradoxical nonphysical
nature'' of the model and clarified some misunderstanding surrounding it.
We have not found Pitowsky explicitly acknowledging this
form, but it is part of his treatment of the
Bell-KS Theorem in~\citep{pitowsky1983,pitowsky1989}.}

Note that Pitowsky's spin \nicefrac12 model
features different notions of probability and certainty: 
By design, $\sigma^1_{\vb*a}=-\sigma^2_{\vb*a}$ always holds. Given $\sigma^1_{\vb*a}=+1$, the correct relative frequency for 
for $\sigma^1_{\vb*b}=+1$ is provided by
classical probability (i.e., Lebesgue measure).
However, the absolute frequency for 
$\sigma^1_{\vb*a}=+1$ 
cannot be provided
by classical probability (as there is no measurable
spin $\nicefrac12$ function), instead  Pitowsky evokes his new notion of  probability.

It turns out that once you agree to use Pitowsky probabilities, and abandon the
use of classical Lebesgue measure 
for the relative frequencies as well, you
get a much simpler model, and furthermore
a ``universal'' model (called Kolmogorovian model by Pitowsky), that applies 
not only to the singlet state, but to
any
quantum mechanical situation. 
This is what Pitowsy does in
\citeyearpar[Sec.~5]{pitowsky1989}.
The new construction does not require the Continuum Hypothesis
anymore (but Pitowsky uses the Axiom of Choice).

In this sense (also in Pitowsky's presentation) the Kolmogorovian model replaces 
the spin \nicefrac12 function model for 
the singlet state, so
we will concentrate on the Kolmogorovian model
in the following.

\subsection{Pitowsky's universal Kolmogorovian model}\label{sec:model2}
In this section, we will present Pitowsky's
\citeyearpar[Sec.~5]{pitowsky1989} 
 ``universal'' Kolmogorovian model.
This model deals with (orthogonal) projection operators only.
I.e., instead of predicting which real numbers
will be the result of a measurement,
we just predict whether a test will result in ``yes''.\footnote{When investigating which kind of 
hidden variables might exist,
the restriction to projections 
does not make a substantial difference.}

\subsubsection*{Kolmogorovian models}

Let us give us Pitowsky's definition of a mixture and a Kolmogorovian model right away:
Fix a $\sigma$-algebra $\Sigma$ of (not necessarily 
Lebesgue measurable) subsets of $[0,1]$.
\begin{itemize}
\item   
A mixture $\rho:\Sigma\to\mathbbm{R}$ is a monotone function 
such that $\rho(X)$ is between the inner and the outer Lebesgue measure of $X$.\footnote{\label{fn:innerouter}Not every $X\subseteq[0,1]$ is
(Lebesgue) measurable (at least assuming the Axiom of Choice);
but for every $X$ there is the inner  measure $m_*(X)$
and the outer  measure $m^*(X)$ with the following property: $m_*(X)$ is the maximal measure of any 
measurable subsets $Y_*$ of $X$; and $m^*(X)$ is the minimal measure of any  measurable $Y^*$ containing $X$, i.e.,
$Y^*\supseteq X$. Of course $m_*(X)\leq m^*(X)$,
and equality holds if and only if $X$ is measurable (and then
the measure of $X$ is $m_*(X)= m^*(X)$). 

If $X_1$ and $X_2$ are disjoint, i.e., $X_1\cap X_2=\emptyset$,
and $X_1$ has outer measure $1$, then $X_2$ has inner measure $0$. To see this, note that
any $Y_*\subseteq X_2$ satisfies $[0,1]\setminus Y_*\supseteq X_1$, so if $Y_*$ had measure $\epsilon>0$ then
$[0,1]\setminus Y_*$ had measure $1-\epsilon<1$.}

\item 
A Kolmogorovian model consists of the following objects, with the following properties:
\begin{itemize}
\item[I] $[0,1]$ is the set of hidden variables. A
hidden variable 
$\lambda\in[0,1]$
maps each 
projection $A$ to  some $\lambda(A)\in\{0,1\}$
(the result of $A$ when measuring the system in the hidden variable state $\lambda$). We set  $X_A:=\{\lambda\in\Omega:\, \lambda(A)=1\}$; and require that 
$X_A\in\Sigma$.
\item[ II]  Given  (a specific method to prepare) a quantum mechanical state $\ket\Psi$, we fix a  mixture $\rho$ (which tells us how the hidden variable states of the state $\Psi$ are distributed; i.e., how likely
a specific variable $\lambda$ will occur).
\item[III] If $A$ and $B$ can be measured simultaneously (i.e., they commute), then
\[
\rho(X_A\cap X_B)+\rho(X_A\cup X_B)=\rho(X_A)+\rho(X_B).
\]
\item[IV]
In a ``random sample'' of copies of the state $\ket\Psi$,  
the relative frequencies of the outcome $1$ for $A$ 
(i.e., for $\lambda\in X_A$)
``approaches'' $\rho(X_A)$ as the sample grows.
\end{itemize}
\end{itemize}

Contrast this definition with the ``statistical  hidden variables'' of Section~\ref{sec:blah}:
Instead of using a ``classical'' probability measure,
$\rho$ is now just a ``mixture''. Also,
for a measure the item corresponding to IV 
is trivially satisfied (it is just a version of the law of large numbers); while now IV is not even well defined\footnote{
Pitowsky mentions that requirement IV is ``not 
a mathematical theorem, it is merely a consistent claim'' \citep[p.~163]{pitowsky1989}. We would not even call it a claim, as the notions are undefined.} (it is not clear what either ``random'' or ``approaches'' actually mean; as we will see later 
``approaches'' cannot mean that a limit in the usual mathematical sense exists).


\subsubsection*{Generalized Kolmogorovian models}\label{sec:genkol}

Instead of directly presenting Pitowsky's  construction
of  a Kolmogorovian model, 
we will first give a simpler (but really equivalent) 
construction for a modified notion.

Let us first note that it is physically irrelevant to require the 
hidden variables to be real numbers (as opposed to, say, vectors in a Hilbert space);
and to require that $\rho(X_A)$ should have anything to do with inner or outer Lebesgue measures.
Dropping these artificial requirements, we get:

\begin{definition} Fix an arbitrary set $\Omega$ (the set of hidden variables) and $\Sigma$, a $\sigma$-algebra of subsets of $\Omega$.
\begin{itemize}
\item   
A \textbf{generalized mixture} $\rho$ is a monotone
function from $\Sigma$ to $\mathbbm{R}$.
\item 
A \textbf{generalized Kolmogorovian model} 
is a Kolmogorovian model that uses $\Omega$  instead of $[0,1]$ as
set of hidden variables and a generalized mixture instead of a mixture.\footnote{More verbosely, the definition is identical to the definition of Kolmogorovian model, where in Item I 
we replace $[0,1]$ with $\Omega$ and in Item II ``mixture'' with ``generalized mixture''.}
\end{itemize}
\end{definition}

We can now (without using the Axiom of Choice or 
non-measurable sets) 
construct a trivial (and, of course, physically useless) generalized Kolmogorovian
model:
We fix a quantum mechanical system with Hilbertspace $\mathcal H$, and a state $\ket\Psi$.
\begin{itemize}
\item We let $\Omega$ be the (pure) quantum mechanical states 
(i.e., elements\footnote{We could instead use one-dimensional subspaces, but that would make the notation slightly less convenient.} of $\mathcal  H$ with norm $1$).
\item 
For a hidden variable $\lambda\in\Omega$ and 
an (orthogonal) projection $A$ (which we identify with its image 
$H_A$), we set $\lambda(A)=1$ if 
$\lambda$ is eigenvector of $A$ with eigenvalue $1$, i.e., if 
$\lambda\in H_A$. (And we set $\lambda(A)=0$ otherwise.) 
Accordingly $X_A=\{\lambda:\, \lambda(A)=1\}$, which is
the set of normalized elements of $H_A$. 

\item
We let $\Sigma$ be the set of all subsets
of $\Omega$.\footnote{Or we could let $\Sigma$ be the $\sigma$-algebra 
generated by the sets $X_A$, it doesn't matter.}

\item
For $Y\in\Sigma$,
let $\langle Y\rangle $  be the 
$\mathcal  H$-subspace 
generated by $Y$.\footnote{In the infinite dimensional case we might want to
use ``closed subspace'' instead of ``subspace''.} 

We set
$\rho(Y)$ to be the quantum mechanical probability
(using the state $\ket\Psi$)
for 
the orthogonal projection to
the subspace $\langle Y\rangle $ resulting in $+1$.

So $\rho(X_A)$ is the quantum mechanical
probability for $A$ resulting in $+1$.
\end{itemize} 
It is easy to check that $\Omega,\Sigma,\rho$ forms
a generalized Kolmogorovian model:

$\rho$ is a generalized mixture as it is monotone.
I and II are obvious.
III is satisfied as well: Assume $A$ and $B$ commute.
Then $\rho(X_A)$,
$\rho(X_B)$, $\rho(X_A\cap X_B)$ and $\rho(X_A\cup X_B)$ are the quantum-mechanically predicted vales for a positive outcome of
$A$, $B$, ``$A$ and $B$'', ``$A$ or $B$'', respectively; and
therefore satisfy the equality in III.\footnote{
In more detail:
Fix an orthonormal basis $\mathcal L$ of common eigenvectors
for $A$ and $B$.
Then $\mathcal L$ is the disjoint union of $\mathcal L_i$
for $i={A\cap B},{A\setminus B},{B\setminus A},
\lnot(A\cup B)$, where
$\mathcal L_{A\cap B}$ consists of $1$-eigenvectors for both $A$ and $B$, $\mathcal L_{A\setminus B}$ consists of vectors that 
are $1$-eigenvectors for $A$ and $0$-eigenvectors for $B$, etc.
Then $\langle X_A\cap X_B\rangle = \langle\mathcal L_{A\cap B}\rangle$,
and thus $\rho(X_A\cap X_B)$ is ``the probability  for $\langle\mathcal L_{A\cap B}\rangle$'', i.e.,
$\Sigma_{v\in \mathcal{L}_{A\cap B}}\abs{\braket{\Psi}{v}}^2$.
Similarly,
$\langle X_A\rangle=X_A=\langle \mathcal L_{A\cap B}\cup \mathcal L_{A\setminus B}\rangle$, so
$\rho(X_A)=\Sigma_{v\in\mathcal L_{A\cap B}\cup \mathcal L_{A\setminus B}}\abs{\braket{\Psi}{v}}^2$.
Analogously for $X_B$.
And $\langle X_A\cup X_B\rangle=\langle
\mathcal L_{A\cap B}\cup \mathcal L_{A\setminus B}\cup
\mathcal L_{B\setminus A}\rangle$.
So $\rho(X_A\cup X_B)=\Sigma_{v\in\mathcal L_{A\cap B}\cup \mathcal L_{A\setminus B}\cup \mathcal L_{B\setminus A}}\abs{\braket{\Psi}{v}}^2$.

Remark: Other than $X_A\cap X_B$, $X_A\cup X_B$ is generally not the set of unit vectors of a subspace of $\mathcal H$.
}

\subsubsection*{From generalized Kolmogorovian models to regular ones}

We now use $\Omega,\Sigma,\rho$ to 
recover Pitowsky's construction of 
a Kolmogorovian model $\Sigma',\rho'$
which satisfies the original definition (which, as we would
like to stress again, has the same physical contents as the generalized notion). For this, we just add a bit of  simple measure theory:

Fist note that the set $\Omega$ of states has size continuum.
Write the interval $[0,1]$ as the disjoint union of continuum many sets
$(Y_i)_{i\in \Omega}$
of outer Lebesgue measure 1. (Here we use the Axiom of Choice.)

We now declare a real $r\in[0,1]$ to be a hidden variable.
Such an $r$ is element of exactly one $Y_\lambda$ for $\lambda\in\Omega$; and we let $r$ produce the same predictions as $\lambda$ in
our generalized Kolmogorovian model $r(A)=\lambda(A)$.

In particular, $X'_A=\{r\in[0,1]:\, r(A)=1\}=\bigcup_{\lambda\in X_A}Y_\lambda$
will contain some of the ``blocks'' $Y_\lambda$
completely, and will omit others completely, more formally:
\begin{equation}\tag{$\ast$}\label{eq:wfdjhw}
\text{If $r\in X'_A$ and $s$ is in the same block $Y_\lambda$ as $r$, then $s\in X'_A$.}
\end{equation}
So we really just replace a single vector $\lambda\in\Omega$
with the block $Y_\lambda$.

Let $\Sigma'$ be the $\sigma$-algebra generated
by the sets $X'_A$. Then every element $X'$ of $\Sigma$
satisfies~\eqref{eq:wfdjhw} as well.\footnote{$\Sigma$
could be defined as closure under arbitrary unions instead of 
just countable ones, or even as the set of all subsets
of $[0,1]$. In the latter case we just have to redefine 
$\rho'(X')$ as $\min\left(m_*(X'),\max\left(m^*(X'),\rho(X)\right)\right)$.}
In particular, whenever $X'\in \Sigma'$ is neither empty
nor equal to $[0,1]$, it has outer measure $1$ and
inner measure $0$ (cf.~the end of
Footnote~\ref{fn:innerouter}).

For $X'\in  \Sigma$, we set $X=\{\lambda\in \Omega:\, Y_\lambda\subseteq X'\}$ and $\rho'(X')=\rho(X)$.
In particular, $\rho'(X'_A)=\rho(X_A)$, which is the
quantum mechanical probability (assuming state $\ket{\Psi}$) for $A$ resulting in $+1$.

$\Sigma',\rho'$ is a Kolmogorovian model:

As $\rho$ is monotone, so is $\rho'$. 
We have seen 
$0=m_*(X')\le \rho'(X')\le m^*(X')=1$
for nontrivial $X'\in\Sigma$, and for the trivial cases
note that 
$\rho'(\emptyset)=0$ and $\rho'([0,1])=1$.
So $\rho'$ is a mixture.
I and II are obvious.
For III, it is enough to note that 
$X'_A\cup X'_B$ consists of the blocks $Y_\lambda$
that satisfy $\lambda\in X_A\cup X_B$, and so
$\rho'(X'_A\cup X'_B)=\rho(X_A\cup X_B)$. The same holds for $\cap$
instead of $\cup$. 
So the required equation in III for $\rho'$ follows
from the equation for $\rho$.

%
%
%
%
%
%
%
%
%
%
%
%

\subsubsection*{A simple example}

Already a very simple example shows the strange nature
of Pitowsky's model:\footnote{An even simpler example
for the same effect
uses just a single spin-$\nicefrac12$ particle;
where we have to use non-contextuality instead of locality.}

Let us look at a pair of spin-$\nicefrac12$ particles in the singlet state. We use the ``generalized'' notation (a hidden variable is a state $\phi$ in the four dimensional Hilbert space), which 
can trivially be translated into Pitowsky's original notation (then the hidden variable is a real $r$
that gets mapped to $\phi$).

The hidden variable $\phi$
will result in 
$\sigma^1_{\vb*a}=+1$ if and only if $\phi=\ket{\vb*a}\otimes v$ for some 
$v\in \mathcal{H}_2$ (the two-dimensional space of the second particle); and will result in 
$\sigma^2_{\vb*a}=+1$ if and only if $\phi=v\otimes \ket{\vb*a}$ for some 
$v\in \mathcal{H}_1$.

Measuring $\sigma^1_{\vb*a}$ will not interfere with
the $\sigma^2_{\vb*a}$ hidden-variable measurement, as the 
hidden variable model is local.\footnote{As explicitly claimed by~\citet[(b) on p. 169]{pitowsky1989}.}

Let us denote with $X_{\vb*a}$ the set of hidden variables that 
have the form either $v\otimes \ket{\vb*a}$ or
$\ket{\vb*a}\otimes v$.
As we can perform both (spatially separated) measurements, and as the Pitowsky-probabilities correspond to the 
quantum mechanically predicted probabilities, 
we know that $X_{\vb*a}$ has Pitowsky-probability 1
(i.e., $\rho(X_{\vb*a})=1$).

So for three different directions
$\vb*a$, $\vb*b$ and $\vb*c$ the sets  $X_{\vb*a}\cap X_{\vb*b}$ have at most two elements while
\begin{equation}\label{singletss}
X_{\vb*a}\cap X_{\vb*b}\cap X_{\vb*c}\text{ is empty.}
\end{equation}
(Really empty, not just of probability 0.)



So, to paraphrase Pitowsky: \emph{All} balls are 
red, and \emph{all} balls are small, and \emph{all}
balls are heavy, but \emph{no}
ball is small and red and heavy.

While this effect is most obvious in the Kolmogorovian model, similar effects
apply to the earlier model
for the singlet state\footnote{This specific example however
does not work there, as
$\sigma^1_{\vb*a}=-\sigma^2_{\vb*a}$ always holds.} due to
Bell's theorem,  as described in
sections Criticism and Response on page~\pageref{sec:criticism} and also in~\citep[p. 1331]{jozsa}, see also~\citep{Czachor1992}.

\section{Analysis of the Kolmogorovian model}

\subsection{Kolmogorovian models are superdeterministic}
\subsubsection*{The weirdness of quantum mechanics}

To work with another example,\footnote{\citet{pitowsky1989} does not mention the GHZ Theorem, 
which was not yet known. However, 
he mentions (and claims that his model deals with)  the Bell-Kochen-Specker Theorem
\citep{Bell1964,kochenspecker},
which is very similar to GHZ but requires not just locality, but non-contextuality of the hidden variables
(which holds in the Kolmogorovian model).}
let us move from the singlet state to the GHZ state.
According to Fact~\ref{thm:GHZ},
quantum physics implies that
each of $\circled1$--$\circled4$ 
is satisfied, and that this cannot be done by fixing hidden
parameter values $\pm1$ for the six $\sigma^i_{\vb*a}$
(for $i\in\{1,2,3\}$ and $\vb*a\in\{x,y\}$).

So all deterministic hidden variables theories have to be ``weird''
in some way or the other. They have
to suffer from one of the following:
\begin{enumerate}[label=(\alph*)]
\item Nonlocality. As mentioned, there are such
non-local hidden variable models.
\item Super-determinism. As mentioned, physically 
doubly irrelevant: Firstly super-determinism
is physically unfeasible, secondly hidden
variables are pointless within super-determinism.
\item Non-classical logic.
One could maybe claim that $\circled1$--$\circled4$
\emph{can} hold simultaneously for suitable
``yes/no'' values, if we completely change
our basic understanding of logic and reasoning.  
We will ignore such positions here.
\end{enumerate}

\subsubsection*{Pitowsky's probability}

Pitowsky however claims to have found another way:
supposedly one can move
all the weirdness into his nonstandard notion
of \emph{statistics} (Pitowsky probability). He explicitly claims~\citep[p. 169]{pitowsky1989} that his model is local (even non-contextual) and classical, i.e., does not suffer weirdness (a) or (c).
We will argue in the following that Pitowsky's model is in fact just super-deterministic.

Let $A_1$ be the set of hidden Kolmogorovian 
variables satisfying $\circled{1}$.
As quantum mechanics implies $\circled{1}$,
$\rho(A_1)=1$. In other words, 
the Pitowsky probability for a hidden variable to be in the set $A_1$ is 1.
Analogously define $A_2$, $A_3$ and $A_4$.
So $A_i$ all are Pitowsky measure 1 sets; while the intersection
$\bigcap_{i=1,\dots,4}A_i=\emptyset$
is empty.

For Pitowsky, this is a peculiar property
of his notion of probability:
$\circled1$ is satisfied with measure 1 (i.e., always) as $A_i$ has measure 1; the same 
holds for $\circled2$ etc. It is true that 
no hidden variable can have all properties 
$\circled1$--$\circled4$, but if we 
measure property $\circled1$ on some sequence, 
then $\circled1$ will be satisfied;
and $\circled2$ has to be measured on another
sequence so consistently we can assume that
on this sequence $\circled2$ is satisfied, etc.

While Pitowsky claims that this can be seen as
an
effect of a non-classical notion of probability, 
it really is just a variant of super-determinism.
This can be seen most clearly if we just 
consider a single state three-particle system,
which we \emph{first} assume to have hidden variable $v$. We know that $v$ determines 
the six values used in the 
requirements $\circled1$--$\circled4$,
and therefore we know that at least one of the 
requirements fail. Let us assume that 
$\circled1$ fails. In other words:
If we test $\circled1$, we will get a contradiction to
quantum mechanics. However, $\circled1$
will hold with Pitowsky probability 1,
i.e., always.
So in the Pitowsky model, we \emph{will not} measure 
$\circled1$, i.e., our hidden variable excludes 
certain measurements, i.e., is super-deterministic.

\subsubsection*{Isn't a single system unrealistic?}

One could object to the use of a single three-particle 
system: In real world experiments, one has to deal
with non-perfect measurements etc., and a violation
of quantum mechanic predictions in a single event
would just be considered an outlier and ignored, quite consistently with 
experimental data.

Of course, this argument does not help:
Assume that we \emph{first} create a sequence
of a million hidden variables.
Each of the variables will violate at least 
one of the requirements $\circled1$--$\circled4$; 
so at least one requirement has to be violated
in at least 25\% of the variables.
Again, assume that is the case for $\circled1$.
If the Pitowsky setup were non super-deterministic, we could now choose to test
$\circled1$ on the sequence, and get a 
success rate of at most 75\% (instead
of the 100\% predicted by quantum mechanics); 
which is definitely not in line with real life experiments. 

Two remarks:
\begin{itemize}
\item
If each time we choose the one forbidden setting, 
we of course even get 100\% failure. 
But it seems hard to argue that 
we could guess a long sequence of forbidden settings correctly.
\item The argument relies on
locality only; it is not required 
that the choice of a measurement at position 1, say, will not affect the 
outcomes for subsequent measurements at the same position 1.
(However, in Pitowsky's 
Model the outcome is in fact independent).
So we generally
have to assume that the result for, e.g., the $n$-th 
measurement of $\sigma^1_x$ also depends on the
measurement setting ($x$, $y$ or $z$) of all the previous $n-1$ many
measurements at position 1 (and the measurement results,
but they are determined anyway).
This does not change the argument: 
Using only locality, we can still argue that
the hidden variables will give for given $n$, $i$, and $\vb*a$ 
the same results
for testing 
$\sigma^i_{\vb* a}$
on the $n$-th particle for both appearances
in the equations $\circled1$--$\circled4$.
\end{itemize}

\subsubsection*{The singlet state}

In the GHZ case, we have seen that 
at least one out of four measurement combinations 
will be ruled out by the hidden variable.
In the singlet state, super-deterministic nature of
the Kolmogorovian model
is even more dramatic.


Let us \emph{first} create a 
hidden variable state.
Using the notation of~\eqref{singletss}, this
state can be in $X_{\vb*b}$
(the states that are ``OK'' for $\vb*b$)
for at most two directions, $\vb*b_1 $ and $\vb*b_2$, say.
So whenever we \emph{afterwards} 
chose to measure $\sigma^1_{\vb*c}$ and
$\sigma^2_{\vb*c}$ for \emph{any} direction $\vb*c$ other than
 $\vb*b_1 $ and $\vb*b_2$,
we get ``not $+1$'' both times, 
in a violation of quantum mechanics.

So  if we 
have a device that 
can measure spins in $\ell$ many directions, and if we 
use two of these devices, spatially separated,  and make sure both will measure the 
same direction,\footnote{E.g., both devices look at the same distant star and use photons arriving from there to 
determines the direction in a previously agreed way.}
then for any given hidden variable state, at least
$\ell-2$ out of the $\ell$ possible measurements will violate quantum mechanics.

\subsection{From super-determinism to the Kolmogorovian model, Pitowsky's law of large numbers}

We have argued that Kolmogorovian models imply
a form of super-determinism. 

In turn,
from a super-deterministic point of view,
we can easily 
see how the Kolmogorovian model works. This will
also shed light on how to interpret the notions in assumption IV of the definition of the Kolmogorovian model:
\begin{quote}
``In a random sample of particles, whose 
statistical state is given by the mixture $\rho$,
the frequency of particles having property $A$
approaches $\rho(A)$ as the sample grows.''~\citep[p.~163]{pitowsky1989}
\end{quote}
Recall that $\rho(A)$ is just the quantum mechanical probability for $A$.

So let us assume a super-deterministic position.
We know (since we know all) that we will investigate a sequence of $m$
many systems (in state $\ket\Psi$),
and that for each copy of the system we
will measure the sequence
$A_1,A_2,\dots,A_n$ of commuting\footnote{As mentioned, whenever we measure
a non-commuting observable, we will have to change
any hidden parameter.}
projections. We know that the 
$j$-th system will give the 
results $a^j_1,\dots,a^j_n$.
If $m$ is large enough, then (with high probability)
the $a^j_i$ will appear with frequencies close to the quantum mechanically predicted sequences.
For each $j$, 
there is a state $\phi^j$ which is simultaneous
eigenvalue of $A_i$ with eigenvalue $a^j_i$.
This is the $j$-th hidden variable we chose.

So in this construction, where we start from super-determinism,
the notions in assumption IV are the following:
``random'' means according to our prescience (i.e., quite
determined and not random at all). 
``Approaches'' means the usual limit, but only for
those $A$ that are actually measured. (Other observables $A$ will feature frequencies entirely different from $\rho(A)$.)\footnote{We could
also choose the following alternative (but this is not what
happens in the Kolmogorovian model): We could,
from the pre-determined sequence of hidden variables, 
define $ \rho(A)$ to be the frequency for $A$
as determined by the hidden variables.
In that case, ``approaches'' will 
be the limit for all $A$, but the value of $\rho(A)$
will differ
from quantum mechanical predictions (in those observables $A$ that
are not actually measured.}

In this sense, Pitowsky's model is logically consistent (just as super-determinism is). Of course, this is not particularly
satisfying; but as we have argued over and over, we cannot do better. 

Let us argue the same thing once again, from a slightly different point of view:

Classically, we have the following (which can
be seen as a simple form of the law of large numbers):
\begin{quote}
Fix finitely (or countably)  many Lebesgue measurable sets  $A_i\subseteq [0,1]$ of measure $a_i$. 
Generate a random\footnote{Using the ``standard'' notion of probability: We use independent and identically distributed random variables, each using the uniform distribution. 
In other words,
we use the product measure of 
countably many copies of the Lebesgue measure on $[0,1]$.}
sequence $(r_n)_{n\in\mathbb N}$ 
of elements of $[0,1]$.
 Then with probability 1, 
the frequency of $r_n\in A_i$ is $a_i$
for all $i$.\footnote{I.e., $\lim_{n\to\infty}\frac{|\{1\le m\le n:\, r_m\in A_i\}|}{n}$ exists and is equal to $a_i$.}
\end{quote}
Pitowsky observes that he cannot prove a
similar law for his notion of probability;
and replaces it with Principle IV in his definition of Kolmogorovian models, quoted above.

It is true that a ``proper'' Pitowsky law of large numbers would be sufficient
to give a realistic, local, non-super-deterministic hidden variable theory;
but such speculation is vacuous as 
it is easy to see that the ``proper'' law fails 
for Pitowsky probability.
Let us again use the GHZ notation:

\begin{quote}
Fix sets $A_i\subset[0,1]$ ($i\in\{1,2,3,4\}$)
such that $\bigcap A_i=\emptyset$.
They do not have to be Lebesgue measurable (but we 
will assume they are ``Pitowsky measurable'' and all have 
``Pitowsky measure'' 1).
Generate (by whatever process you like, e.g.,
a Pitowsky random process) an infinite sequence $(r_n)_{n\in\mathbb N}$ of elements of $[0,1]$.
Then for at least one $i\in\{1,2,3,4\}$, the 
frequency of the property $r_n\in A_i$ is not 1. 

Actually,
if all frequencies exist\footnote{I.e., 
if $f_i:=\lim_{n\to\infty}\frac{|\{1\le m\le n:\, r_m\in A_i\}|}{n}$ is defined}
then for at least one $i$ the frequency is  $\le0.75$.
\end{quote}

The proof is entirely trivial: 
Fix any sequence $r_n$, and assume that all
frequencies are defined and bigger than $0.75$. I.e., for each $i\in\{1,2,3,4\}$ there is a $N_i$ such that 
for all $n\ge N_i$ the $A_i$-frequency of the finite
sequence $r_1,\dots,r_n$
is bigger than $0.75$.
Let $n$ be $\max(N_1,N_2,N_3,N_4)$.
Then the finite sequence up to $N$ 
has frequency bigger than $0.75$ for all $i$, a contradiction of the fact that $\bigcap A_i=\emptyset$.

Note that our observations in this section do not contradict anything
Pitowsky says directly. Rather,
Pitowsky argues that we 
measure the different $A_i$ on different samples,
and that on each sample we will get the
correct frequency for the according measurement.
(As already mentioned, this is
logically consistent, as it
just means that we are dealing with super-determinism.)

\subsection{Set theory and physics}

There has been some debate over the following question:
\begin{quote}
Could the choice of
set theoretic axioms have any effect on the
physical theories? Or, from a different perspective:
could physical knowledge imply that specific 
set theoretic axioms should be preferred over others?
\end{quote}
Of course we know that, e.g., the Axiom of Choice
is required for many 
mathematical theorems (such as: every
vector space has a basis), which in turn can be 
applied in physics. However, on closer inspection it
turns out that for all concrete instances that are used, the axiom of choice is not required.
The same applies even for the existence of an infinite 
set: One can use a very constructive, 
``finitary'' form of mathematics that is perfectly 
sufficient for physics.
(It is a different question whether it is equally
practical and intuitive as the ``usual'' mathematics
based on set theory.)

Pitowsky used the Continuum Hypothesis
to construct 
a spin-$\nicefrac12$-function model. 
Pitowsky suggested that the existence of such a function
might not follow
from the usual axioms of set theory alone,
which has 
recently been confirmed by~\citet{farahmagidor}.
In the same paper, as well as in~\citep{magidor},
it has been argued that the spin-$\nicefrac12$ 
model is an 
indication 
that physical considerations
might provide input on which new axioms
should be adopted for set theory.

We do not share this opinion:
\begin{itemize}
\item Pitowsky's spin-$\nicefrac12$-function 
is introduced just to ``sabotage'' a specific 
argument in one of the possible proofs of Bell's theorem: Using
a non-measurable function prevents proofs using the
expected value of the function. Accordingly, 
the spin-$\nicefrac12$-function alone does not give
any model; the real point of the matter is the 
dramatically altered basic notion of probability
which has to be added on top of it (let us call it
Pitowsky probability).
\item Once you adopt Pitowsky probability, the spin-$\nicefrac12$-function
becomes unnecessary: It is easy to give a universal 
model for all quantum mechanical systems (the Kolmogorovian model). Pitowsky uses the Axiom of Choice  for this model, but we think even that is unnecessary\footnote{We may need some implications of the Axiom of Choice that allow us to work with the Hilbertspace in the usual way; but we do not need the partition of the reals into continuum many disjoint unmeasurable sets.}, as argued in the section on generalized Kolmogorovian models in Section~\ref{sec:model2}.
\item But in the end, Pitowsky probability turns out
to be just a variant of super-determinism.
Accordingly, the models are obviously consistent, but physically not relevant. (And doubly so: 
super-determinism is physically unfeasible,
and hidden
variables are pointless within super-determinism.)
\end{itemize}

So we come to quite the opposite conclusion as Farah and Magidor: Instead of indicating connections between physics and set theory, Pitowsky's attempts of hidden variables rather seems to 
reaffirm the old intuition: ``if nontrivial set theory, non-constructive mathematics or a non-measurable set is used in
an essential way, it cannot be physically relevant''.

\bibliography{SuperPitowsky}
\end{document}